\documentclass[letterpaper]{SWARM19}
\usepackage{amsmath,amsfonts,amssymb}
\usepackage{graphicx}
\usepackage{subcaption}
\usepackage[super]{nth}
\newtheorem{dfn}{\bf Definition}

\newtheorem{exa}[dfn]{\bf Example}
\title{Distributed Possibilistic Learning in Multi-Agent Systems}

\author{Jonathan Lawry${}^{1}$, Michael Crosscombe${}^{1\dagger}$ and David Harvey${}^{2}$ }
\speaker{Michael Crosscombe}

\affils{${}^{1}$Department of Engineering Mathematics, University of Bristol, BS8 1UB, UK\\
${}^{2}$Research Technology and Innovation, Thales UK}
\abstract{%
Possibility theory is proposed as an uncertainty representation framework for distributed learning in multi-agent systems and robot swarms. In particular, we investigate its application to the best-of-$n$ problem where the aim is for a population of agents to identify the highest quality out of $n$ options through local interactions between individuals and limited direct feedback from the environment. In this context we claim that possibility theory provides efficient mechanisms by which an agent can learn about the state of the world, and which can allow them to handle inconsistencies between what they and others believe by varying the level of imprecision of their own beliefs. We introduce a discrete time model of a population of agents applying possibility theory to the best-of-$n$ problem. Simulation experiments are then used to investigate the accuracy of possibility theory in this context as well as its robustness to noise under varying amounts of direct evidence. Finally, we compare possibility theory in this context with a similar probabilistic approach.
}


\begin{document}

\maketitle


\blfootnote{This work was presented at the 3rd International Symposium on Swarm Behavior and Bio-Inspired Robotics (SWARM 2019), Okinawa, Japan, November 16--22, 2019.}

\section{Introduction}
There is growing evidence that providing agents with a way of explicitly representing uncertainty in their beliefs can help to facilitate robust distributed learning for multi-agent systems operating in noisy environments. For example,~\cite{crosscombe} shows that robot swarms which use a third truth-value to represent ``unknown" when solving best-of-$n$ problems~\cite{valentini} are more robust to noise and malfunction than those using binary belief models. In a similar context,~\cite{lawry2} investigates the use of epistemic sets to represent an agents' belief, corresponding to the set of states of the world which they deem to be possible. Extensive multi-agent simulations were then employed to test the scalability of such an approach when applied to large state spaces and varying numbers of agents. Here we extend this idea by introducing possibility theory~\cite{zadeh,dubois2} to capture graded possibilities, whereby instead of simply classifying states as being either possible or impossible, an agent can allocated a degree of possibility between $0$ and $1$ to any given state. In particular, we explore the application of possibility theory to distributed multi-agent learning and more specifically to the best-of-$n$ problem in which a population of agents aims to identify the best out of $n$ possible choices (states) on the basis of feedback from the environment and local interactions between individuals. 

The best-of-$n$ is a class of distributed learning and decision making problems with clear relevance to a range of applications including search and rescue~\cite{peleg}, pollution treatment swarms and distributed task allocation~\cite{kakalis}. Common approaches to best-of-$n$ in swarm robotics are often biologically inspired, for example by the behaviour of honeybees and Temnothorax ants when searching for nest sites~\cite{burns,valentini0,reina}. However, it can also be tackled by using a combination of belief fusion between agents together with evidential updating by individuals based on, for example, sensor readings. The epistemological importance of this approach to distributed learning has been highlighted by~\cite{douven} and has been studied using probabilistic models of belief in~\cite{lee}. In the sequel we employ this combined approach in a possibility theory setting. Furthermore, in this context we will argue that the combination of fusion and evidential updating is more effective and robust to noise than evidential updating alone. 

Possibility theory~\cite{dubois} is a computationally efficient framework for reasoning about uncertainty which allows for a distinction to be made between uncertainty and ignorance. At a representational level there are clear analogies with probability theory. For instance, subjective belief is represented by a possibility distribution which allocates a degree of possibility between $0$ and $1$ to each state of the world. The calculus for possibilities is maxitive in contrast to that for probabilities which is additive. In particular, the degree of possibility of a proposition, as represented by a set of states, is given by the maximum of the possibility values of those states. Possibility theory can also be understood as a special case of Dempster-Shafer theory~\cite{dubois} and consequently also of imprecise probability theory. From the latter perspective, possibility theory provides both an upper bound (possibility) and a lower bound (necessity) on the probability of any proposition, with the difference between these providing a measure of ignorance. This capacity to quantify ignorance then provides additional flexibility when fusing highly inconsistent beliefs, whereby any such combination will lead to an increase in overall ignorance. In the following we will exploit these properties for distributed learning to help facilitate consensus across a population of agents. 

An outline of the remainder of the paper is as follows. Section~\ref{sec:possibility} gives an overview of the basic concepts from possibility theory including a proposed approach to fusing possibility distributions. In Section~\ref{sec:agentbased} we described a discrete time agent-based model for possibility theory applied to the best-of-$n$ problem. Simulation experiment results are then described for this model at different levels of noise and for different evidence rates. In this context the possibilistic approach is compared with a probabilistic approach similar to that outlined in~\cite{lee}. Finally, Section~\ref{sec:conclusions} gives some conclusions.

\section{Possibility Theory}
\label{sec:possibility}
In this section we give a brief introduction to possibility theory and propose a parameterised family of fusion operators as an effective method for combining partially inconsistent beliefs. The underpinning concept is that of a \emph{possibility distribution} which allocates a degree of possibility to each of a set of possible states of the world, denoted by
${\mathbb S}=\{s_1,\ldots,s_n\}$, as follows: \\
\begin{dfn}{Possibility Distribution}\\
\label{dfn:possdist}
\textit{A possibility distribution on ${\mathbb S}$ is a function $\pi:{\mathbb S}\rightarrow [0,1]$ such that $\max\{\pi(s):s \in {\mathbb S}\}=1$. Let ${\mathbb D}$ denote the set of all possibility distributions on $\mathbb{S}$.} \\
\end{dfn}
A possibility distribution characterises a possibility and necessity measure defined over propositions as represented by sets of states. Intuitively, in a context in which knowledge is graded, the possibility measure of a proposition quantifies the degree to which it is consistent with an agent's knowledge, while the necessity quantifies the degree to which it is entailed by that knowledge.\\
\begin{dfn}{Possibility and Necessity Measures}\\
\label{dfn:possmeasure}
\textit{A possibility distribution $\pi$ on ${\mathbb S}$ generates a possibility and a necessity measure on $2^{{\mathbb S}}$ such that, for $A \subseteq {\mathbb S}$,
\begin{align*}
    \Pi(A) &= \max\{\pi(s):s \in A \} \ \text{ and } \\
    N(A) &= \min\{1-\pi(s):s \in A^c \}.
\end{align*}
For notational simplicity we use $N(s_i)$ as shorthand for $N(\{s_i\})$.} \\
\end{dfn}
The following properties of possibility and necessity measures are immediate consequences of Definition~\ref{dfn:possmeasure}~\cite{dubois}:
For $A,B \subseteq \mathbb{S}$,
\begin{itemize}
    \item $\Pi({\mathbb S})=N({\mathbb S})=1$ and $\Pi(\emptyset)=N(\emptyset)=0$,
    \item $N(A)\leq \Pi(A)$ and $N(A)=1-\Pi(A^c)$,
    \item $N(A \cap B)=\min(N(A),N(B))$ and $\Pi(A \cup B)=\max(N(A),N(B))$.
\end{itemize}
Possibility and necessity measures can also be interpreted as upper and lower probabilities respectively in the sense that there is a set of probability measures on propositions satisfying $N(A) \leq P(A) \leq \Pi(A)$ for all $A \subseteq {\mathbb S}$, and furthermore that $N(A)$ and $\Pi(A)$ are respectively the infimum and supremum of this set. The level of ignorance about a proposition $A$ is then given by $\Pi(A)-N(A)$.   

We now propose a family of fusion operators for combining possibility distributions which in the current context represent the beliefs of different agents in the population. This will require the notion of an intersection function, or \emph{t-norm}, as given below. The intuition here is that if we consider the possibility distributions of two agents as defining two fuzzy sets  of possible states~\cite{zadeh} then one approach to fusing them is to take the intersection of these two sets~\cite{dubois3}.\\ 

\begin{dfn}{Intersection Function (t-norm)}\\
\label{dfn:tnorm}
\textit{An intersection function (t-norm) \cite{klir} is a function \\$T:[0,1]^2 \rightarrow [0,1]$ which satisfies the following properties:
\begin{itemize}
    \item  For $x \in [0,1]$, $T(x,1)=x$.
    \item  For $x,y,z \in [0,1]$, if $y \leq z$ then $T(x,y) \leq T(x,z)$.
    \item  For $x,y \in [0,1]$, $T(x,y)=T(y,x)$.
    \item  For $x,y,z \in [0,1]$, $T(x,T(y,z))=T(T(x,y),z)$.\\
\end{itemize}}
\end{dfn}

\begin{dfn}{Possibility Fusion Function}\\
\label{dfn:pool}
\textit{A possibility fusion function is a function \\$c: {\mathbb D}^2 \rightarrow {\mathbb D}$. In particular, we propose the following family of t-norm based pooling functions:
\begin{align*}
    c(\pi_1(s),\pi_2(s)) &= T(\pi_1(s),\pi_2(s))\\
    &+ 1-\max\{T(\pi_1(s),\pi_2(s)):s \in \mathbb{S}\}.
\end{align*}}
\end{dfn}
\vspace{1em}
Definition~\ref{dfn:pool} takes the fusion of two possibility distributions to be their intersection, normalised so that the maximum possibility of the fused possibility distribution is $1$ as required by Definition~\ref{dfn:possdist}. The effect of the normalisation is to increase the level of ignorance if the two possibility distributions are highly inconsistent. We can argue that $\max\{T(\pi_1(s),\pi_2(s)): s \in {\mathbb S} \}$ provides a good measure of the consistency of two possibility distributions $\pi_1$ and $\pi_2$. Intuitively, $\pi_1$ and $\pi_2$ might be considered highly inconsistent if, for all states, $\pi_1$ allocates a high possibility values whenever $\pi_2$ allocates a low possibility value and vice versa. In this case $T(\pi_1(s),\pi_2(s))$ will be low for all states since it follows from Definition~\ref{dfn:tnorm} that  $T(\pi_1(s),\pi_2(s))\leq \min(\pi_1(s),\pi_2(s))$~\cite{klir}. Consequently, the normalising term in Definition~\ref{dfn:pool} will be high as will then be the possibility values of all states. Hence, according to Definition~\ref{dfn:possmeasure} in this case the possibility measure of all propositions, i.e. sets of states, will be close to $1$ and its necessity measure will be close $0$. We now introduce Frank's family of t-norms which generate a parameterised family of fusion operators when used in Definition~\ref{dfn:pool}.
 \\
\begin{dfn}{Frank's Family of t-norms~\cite{frank}}\\
\label{dfn:frank}
\textit{Frank's t-norms are a parameterised family of intersection functions of the form: For $\theta \in {\mathbb R}-\{0\}$
\begin{gather*}
T_\theta(x,y)=-\frac{1}{\theta}\ln\left( 1+ \frac{(e^{-\theta x}-1)(e^{-\theta y}-1)}{(e^{-\theta}-1)} \right).
\end{gather*}
Note that
\begin{align*}
\lim_{\theta \rightarrow 0} T_\theta(x,y) &= x\times y,\\
\lim_{\theta \rightarrow \infty} T_\theta(x,y) &= \min(x,y),\\
\lim_{\theta \rightarrow -\infty} T_\theta(x,y) &= \max(0,x+y-1).
\end{align*}
Furthermore, Frank's t-norms are increasing with $\theta$ such that $\theta \geq \theta^\prime$ implies $T_\theta \geq T_{\theta^\prime}$.}\\
\end{dfn}
In the case that an agent with possibility distribution $\pi$ needs to make a choice, for example by selecting a particular state to investigate further, then one approach is to sample from the pignistic distribution for $\pi$~\cite{smets}; the latter is a member of the set of probability distributions bounded above and below by $\Pi$ and $N$ respectively. Furthermore, it is arguably the least biased such distribution since it allocates probability proportionately to the difference between consecutive possibility values~\cite{smets}. \\

\begin{dfn}{Pignistic Distribution}\\
\label{dfn:pignistic}
\textit{Let $\pi$ be a possibility distribution on $\mathbb{S}$ and let the states be sorted so that $\pi(s_{i+1}) \leq \pi(s_i)$, then the pignistic distribution for $\pi$ is the probability distribution on $\mathbb{S}$ given by
\begin{gather*}
p_\pi(s_i)=\sum_{j=i}^n \frac{ \pi(s_j)-\pi(s_{j+1}) }{j}.
\end{gather*}
Note that the pignistic distribution is order preserving in the sense that $\pi(s_i) \geq \pi(s_j)$ implies $p_\pi(s_i) \geq p_\pi(s_j)$.}\\
\end{dfn}

\begin{figure}[t]
\centering
 \includegraphics[width=0.35\textwidth]{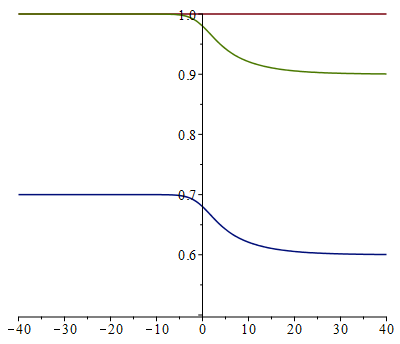}
 \put(-20,130){$c(\pi,\pi^\prime)(s_2)$}
 \put(-20,105){$c(\pi,\pi^\prime)(s_3)$}
 \put(-20,27){$c(\pi,\pi^\prime)(s_1)$}
 \put(-120,-5){\footnotesize Frank's parameter $\theta$}
 \caption{The fused possibility distribution values for $c(\pi,\pi^\prime)$ plotted against the Frank's parameter $\theta$.}
 \label{fig:fusion}
\end{figure}
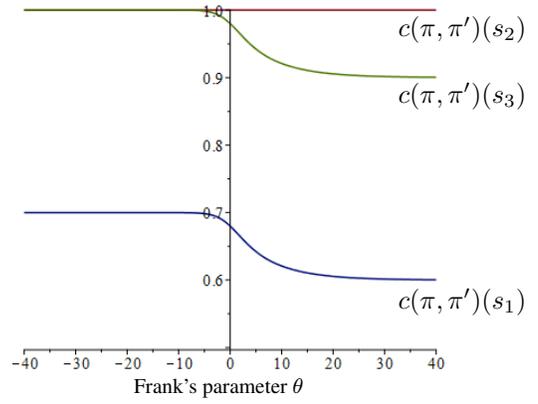

\begin{exa}
\textit{Let ${\mathbb S}=\{s_1,s_2,s_3\}$ and consider the possibility distribution $\pi$ such that $\pi(s_1)=1$, $\pi(s_2)=0.8$ and $\pi(s_3)=0.7$. This generates possibility and necessity measures as given in Definition~\ref{dfn:possmeasure} such that
\begin{alignat*}{3}
&\Pi(\{s_1,s_2,s_3\}) &&= \max(1,0.8,0.7) &&= 1\\
&\Pi(\{s_1,s_2\}) &&= \max(1,0.8) &&= 1,\\
&\Pi(\{s_1,s_3\}) &&= \max(1,0.7) &&= 1,\\
&\Pi(\{s_2,s_3\}) &&= \max(0.8,0.7) &&= 0.8,\\
&\Pi(s_1) = 1, \hspace{0.8em}
\Pi(&&s_2) = 0.8, \hspace{0.8em}
\Pi(s_3) &&= 0.7,
\end{alignat*}
and
\begin{alignat*}{3}
&N(\{s_1,s_2,s_3\}) &&= 1,&&\\
&N(\{s_1,s_2\}) &&= 1-0.7 &&= 0.3,\\
&N(\{s_1,s_3\}) &&= 1-0.8 &&= 0.2,\\
&N(\{s_2,s_3\}) &&= 1-1 &&= 0,\\
&N(s_1) = \min(1&&-0.8,1-0.7) &&= 0.2,\\
&N(s_2) = \min(1&&-1,1-0,7) &&= 0,\\
&N(s_3) = \min(1&&-1,1-0.8) &&= 0.
\end{alignat*}
Hence, for the possibility distribution $\pi$ the degree of ignorance associated with the states $s_1$, $s_2$ and $s_3$ is
\begin{alignat*}{2}
\Pi(s_1)-N(s_1) &= 1-0.2 \ &= 0.8,\\
\Pi(s_2)-N(s_2) &= 0.8-0 &= 0.8,\\
\Pi(s_3)-N(s_3) &= 0.7-0 &= 0.7,
\end{alignat*}
respectively.
The pignistic distribution for $\pi$ is generated according to Definition~\ref{dfn:pignistic} as follows:
\begin{gather*}
p_\pi(s_3)=\frac{0.7}{3}=0.2333,\\ p_\pi(s_2)=\frac{0.8-0.7}{2}+\frac{0.7}{3}=0.2833,\\
p_\pi(s_1)=\frac{1-0.8}{1}+\frac{0.8-0.7}{2}+\frac{0.7}{3}=0.4833.
\end{gather*}
Now consider a second possibility distribution $\pi^\prime$ such that $\pi^\prime(s_1)=0.4, \pi^\prime(s_2)=0.9, \pi^\prime(s_3)=1$. We can fuse $\pi$ and $\pi^\prime$ by applying the operator in Definition~\ref{dfn:pool}. Taking $\theta=10$ and applying the Frank's t-norm as in Definition~\ref{dfn:frank} we obtain
\begin{alignat*}{2}
T_\theta(\pi(s_1),\pi^\prime(s_1)) \ &= \ T_\theta(1,0.4) \ &&= \ 0.4,\\
T_\theta(\pi(s_2),\pi^\prime(s_2)) \ &= \ T_\theta(0.8,0.9) \ &&= \ 0.7791,\\
T_\theta(\pi(s_3),\pi^\prime(s_2)) \ &= \ T_\theta(0.7,1) \ &&= \ 0.7.
\end{alignat*}
In this case the normalising term is \\$1-\max(\{0.4,0.7791,0.7\})=1-0.7791=0.2209$. Hence, the fused possibility distribution is such that
\begin{alignat*}{2}
c(\pi,\pi^\prime)(s_1) \ &= \ 0.4+0.2209 \ &&= \ 0.6209,\\
c(\pi,\pi^\prime)(s_2) \ &= \ 0.7791+0.2209 \ &&= \ 1,\\
c(\pi,\pi^\prime)(s_3) \ &= \ 0.7+0.2209 \ &&= \ 0.9209.
\end{alignat*}
Figure~\ref{fig:fusion} shows the possibility distribution values $c(\pi,\pi^\prime)$ for varying values of $\theta$. In this example the normalisation term is $1-T_\theta(\pi(s_2),\pi^\prime(s_2))$. Since $T_\theta$ is increasing with $\theta$ then this term decreases with increasing $\theta$. In other words, the amount of inconsistency generated by the fusion operator decreases as $\theta$ increases.
}
\end{exa}

\section{Agent-Based Experiments}
\label{sec:agentbased}
We adopt a discrete time agent-based model in order to study the macro-level convergence properties
of a population of agents each attempting to individually solve a best-of-$n$ problem formulated
using possibility theory. Let ${\mathbb A}=\{a_1, \ldots, a_k\}$ denote the set of agents in the population. Each state $s_i \in \mathbb{S}$ has a quality value $q_i \in [0,1]$ which can be sampled with noise from the environment. For example, in a distributed search and rescue scenario, $s_i$ might correspond to a particular location and $q_i$ to the degree of support for there being casualties at that location. We assume without loss of generality that $q_i < q_{i+1}$ for $i=1, \ldots, n-1$. More specifically, in the experiments described below we take $q_i=\frac{i}{n+1}$ for $i=1, \ldots, n$, so that quality values are distributed uniformly across the interval $[0,1]$.

\begin{figure*}[t]
    \centering
    \hspace{2em}
    \begin{subfigure}[b]{0.4\textwidth}
        \includegraphics[width=\textwidth]{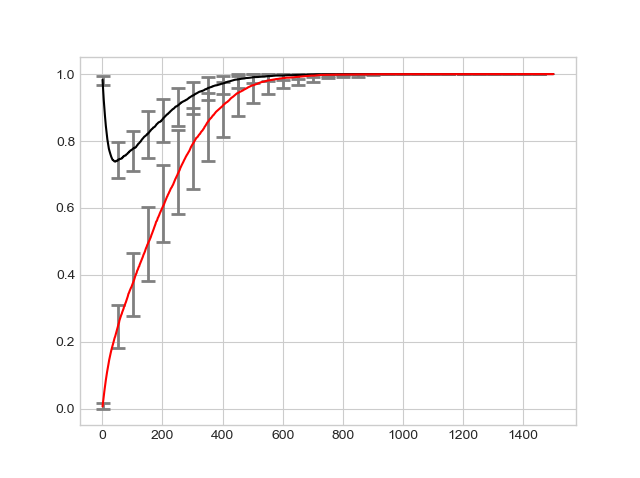}
        \put(-100,0){\footnotesize Time $t$}
        \caption{$\pi(s_n)$ and $N(s_n)$ for fusion and evidential updating.}
        \label{fig:trajectoryboth}
    \end{subfigure}
    \hfill
    \begin{subfigure}[b]{0.4\textwidth}
        \includegraphics[width=\textwidth]{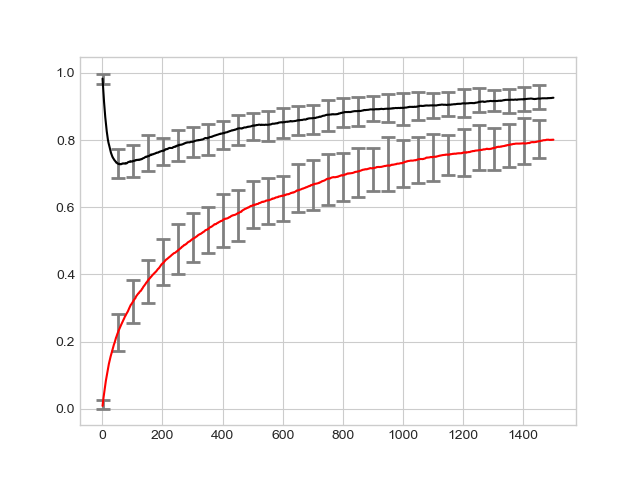}
         \put(-100,0){\footnotesize Time $t$}
         \caption{$\pi(s_n)$ and $N(s_n)$ for evidential updating only.}
        \label{fig:trajectoryevonly}
    \end{subfigure}
    \hspace{2em}
    \caption{Average $\pi(s_n)$ (black line) and $N(s_n)$ (red line) against time for $k=100$, $n=5$, $\rho=0.05$, $\sigma=0$ and $\theta=20$.}
     \label{fig:trajectory}
\end{figure*}

Agents' beliefs about which is the best or highest quality state are represented by possibility distributions on ${\mathbb S}$, and at time $t=0$ all agents in ${\mathbb A}$ are initialised as being completely ignorant so that $\pi(s_i)=1$ for all $s_i \in {\mathbb S}$. At each time step $t$, two agents are selected at random to combine their beliefs by applying the fusion operator given in Definition~\ref{dfn:pool}. The aim here is to model systems with limited communications but mostly unconstrained mixing of agents. The type of swarm robotics or decentralised AI applications on which we are focusing will involve individuals moving independently through an environment and encountering a variety of different agents at different times. This can be modelled by a system where there is free mixing between agents, i.e.\ a totally connected interaction graph of agents, but where there are only relatively few interactions at any given time. The latter is referred to as a ‘well-stirred’ system in~\cite{parker}, corresponding to the assumption that each agent is equally likely to interact with any other agent in the population and that such interactions are independent events. Also, within a time step each agent selects a state to investigate by sampling at random from the pignistic distribution of their current possibility distribution (Definition~\ref{dfn:pignistic}). There is then a probability $\rho$, referred to as the evidence rate, that the agent will succeed in sampling the quality value for their chosen state. $\rho$ provides a simple quantification of the difficulty of obtaining direct evidence from the environment, or of traversing the environment in order to reach the chosen state. If evidence is received then it is with normally distributed noise, so that the agent receives quality $q_i + \epsilon$ for state $s_i$ where $\epsilon$ is a normally distributed random variable with mean $0$ and standard deviation $\sigma$. The agent then updates their belief as follow. The evidence received is represented by a possibility distribution such that\footnote{Since quality values are assumed to be in the range $[0,1]$ then we bound noise values so that the sampled quality is taken to be $0$ if $q_i +\epsilon <0$ and $1$ if $q_i+\epsilon>1$.}
\begin{align*}
    \pi_E(s_j)=
    \begin{cases}
        1 &: j=i,\\
        1-q_i-\epsilon &: \text{otherwise}.
    \end{cases}
\end{align*}
The agent then updates their belief by fusing their current possibility distribution $\pi$ with the evidence to obtain $c(\pi,\pi_E)$.

In summary then we have outlined a discrete time model of a population of agents deployed on the best-of-$n$ problem which is comprised of two decentralised processes both implemented locally by individual agents: evidential updating and belief fusion. The defining parameters of this model are $k$ (the number of agents), $n$ (the number of states), $\rho$ (the evidence rate), $\sigma$ (the noise) and $\theta$ (the Frank's t-norm parameter). In the following subsection we present simulation results for this model for a variety of different parameter values to gain insight into the accuracy and robustness of the possibility theory based approach.

\begin{figure*}
    \centering
    \hspace{2em}
    \begin{subfigure}[b]{0.4\textwidth}
        \includegraphics[width=\textwidth]{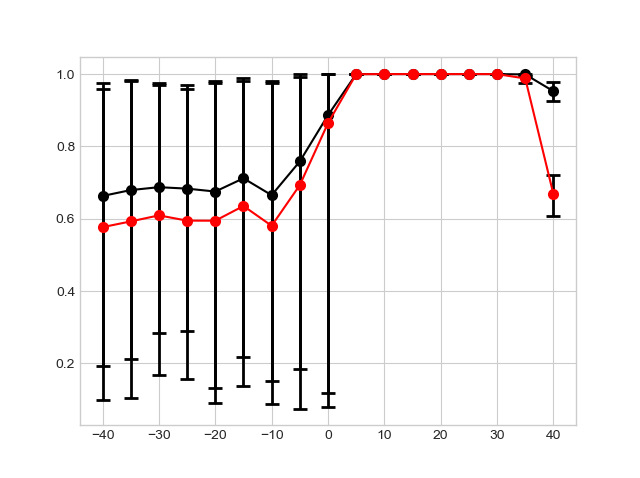}
        \put(-125,0){\footnotesize Frank's parameter $\theta$}
        \caption{$\pi(s_n)$ and $N(s_n)$ for varying $\theta$ and $\sigma=0$.}
        \label{fig:varyings}
    \end{subfigure}
    \hfill
    \begin{subfigure}[b]{0.4\textwidth}
        \includegraphics[width=\textwidth]{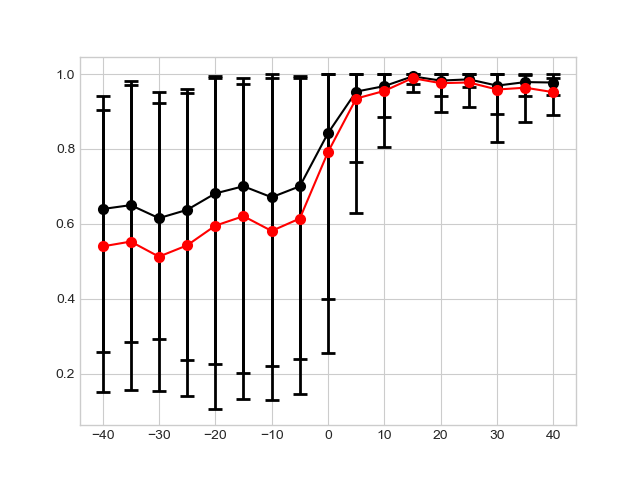}
        \put(-125,0){\footnotesize Frank's parameter $\theta$}
        \caption{$\pi(s_n)$ and $N(s_n)$ for varying $\theta$ and $\sigma=0.3$.}
        \label{fig:varyingsnoise}
    \end{subfigure}
    \hspace{2em}
    \caption{The average values of $\pi(s_n)$ (black line) and $N(s_n)$ (red line) at $t=1500$ for varying $\theta$ both with and without noise for $n=5$ and $\rho=0.05$.}
    \label{fig:dependence}
\end{figure*}

\subsection{Simulation Results}
In the following experiments we simulated the model described above for $k=100$ agents and $n=5$ states while varying the remaining parameters. Each experiment, as characterised by a particular set of parameter values, is run for $1500$ time steps and the results are then averaged over $100$ runs. In all cases averages are shown together with \nth{10} and \nth{90} percentiles represented as error bars. Figure~\ref{fig:trajectory} shows the average possibility and necessity values of the best state, i.e.\ $s_n$, plotted against time. In this case we assume a $5\%$ evidence rate ($\rho = 0.05$), no noise and take $\theta=20$. Figure~\ref{fig:trajectoryboth} shows the results when both fusion and updating take place as described above. In this case the population converges to a shared belief for which $\pi(s_n)=N(s_n)=1$ as characterised by a possibility distribution where $\pi(s_n)=1$ and $\pi(s_i)=0$ for $i\neq n$. In other words, agents are collectively converging on the correct answer with total certainty. Figure~\ref{fig:trajectoryevonly} shows the results when only evidential updating takes place, i.e.\ when there is no fusion of beliefs between agents. Here, convergence is much slower, with the population not having fully converged after $1500$ time steps and agents still maintaining imprecision in their beliefs, as indicated by average necessity values being significantly lower than possibility values. This indicates that fusion can play a positive role in propagating evidence through the population. Furthermore, we will subsequently show that evidence-only learning is much less robust to noise. Initially, however, we consider the sensitivity of the combined model to the value of the Frank's parameter $\theta$.

Figure~\ref{fig:dependence} shows the average values of $\pi(s_n)$ and $N(s_n)$ at time step $1500$ when $\rho=0.05$, plotted against $\theta$ as employed in Definition~\ref{dfn:frank}. Figure~\ref{fig:varyings} is when $\sigma=0.0$, i.e.\ when there is no noise, and Figure~\ref{fig:varyingsnoise} is when $\sigma=0.3$. In both cases we see that performance broadly increases with $\theta$. Note that since Frank's t-norm increases with $\theta$ (Definition~\ref{dfn:frank}), increasing $\theta$ decreases the normalisation term $1-\max\{T_\theta(\pi_1(s),\pi_2(s)):s \in \mathbb{S} \}$ in Definition~\ref{dfn:pool}. In other words, Figure~\ref{fig:varyings} suggests that the population is better at solving the best-of-$n$ when employing fusion operators that do not themselves generate significantly higher levels of inconsistency than is inherent to the two possibility distributions being fused. In light of this, for the remaining experiments we adopt a value of $\theta=20$.

\begin{figure}[b]
    \centering
    \includegraphics[width=0.28\textwidth]{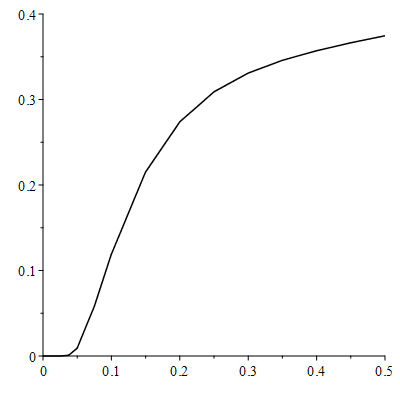}
    \put(-75,0){\footnotesize Noise $\sigma$}
    \caption{Probability that a sampled quality value for $s_5$ will be less than a sampled quality value for $s_4$, plotted against $\sigma$.}
    \label{fig:errorprob}
\end{figure}

\begin{figure}[b]
    \centering
    \begin{subfigure}{0.4\textwidth}
        \includegraphics[width=\textwidth]{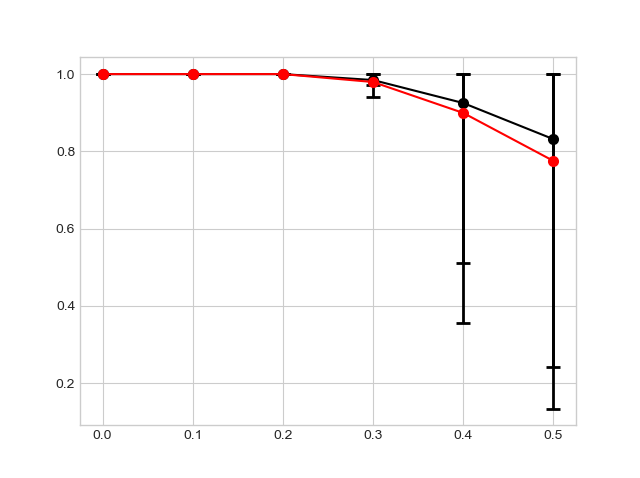}
        \put(-110,0){\footnotesize Noise $\sigma$}
        \caption{$\pi(s_n)$ and $N(s_n)$ for varying $\sigma$ for both fusion and evidential updating}
        \label{fig:varyingnoiseboth}
    \end{subfigure}
    \begin{subfigure}{0.4\textwidth}
        \includegraphics[width=\textwidth]{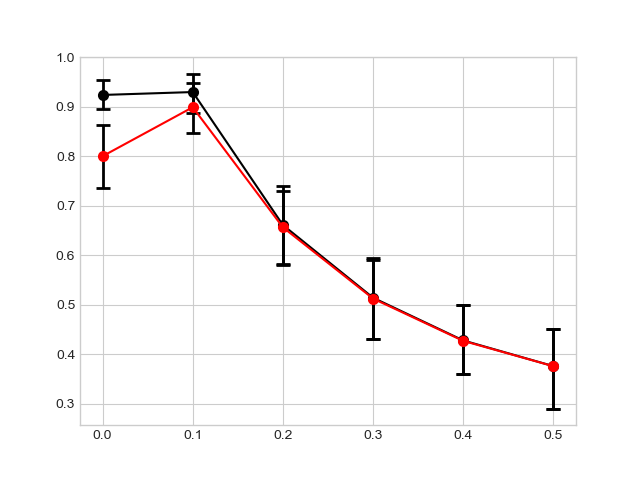}
        \put(-110,0){\footnotesize Noise $\sigma$}
        \caption{$\pi(s_n)$ and $N(s_n)$ for varying $\sigma$ for evidential updating only.}
        \label{fig:varyingnoiseevonly}
    \end{subfigure}
    \caption{The average values of $\pi(s_n)$ (black line) and $N(s_n)$ (red line) after $1500$ iterations for varying levels of noise, when $n=5$, $\theta=20$ and $\rho=0.05$.}
    \label{fig:varyingnoise}
\end{figure}

We now investigate the robustness of the possibilistic approach to noise as represented by the standard deviation of the noise distribution $\sigma$.  In order to give an indication of the effect of noise in this context we can consider the probability that a sampled quality value for $s_5$, the best state when $n=5$, is less than an independently sampled quality value for the second best state $s_4$.  This is shown in Figure~\ref{fig:errorprob} and plotted against the noise standard deviation $\sigma$. For example, when $\sigma=0.3$ the probability that the order of these two quality values will be reversed is $0.331$. Figure~\ref{fig:varyingnoise} shows the average values of $\pi(s_n)$ and $N(s_n)$ at time step $t=1500$, plotted against values of $\sigma$ ranging from $0$ to $0.5$. Figure~\ref{fig:varyingnoiseboth} shows the results when both fusion and evidential updating are combined, while Figure~\ref{fig:varyingnoiseevonly} shows the results when there is only updating from direct evidence. These strongly suggest that combining both fusion between agents and evidential updating results in distributed learning that is much more robust to noise than learning based only on direct evidence. We conjecture that this in part due to the inconsistency handling of the fusion function given in Definition~\ref{dfn:pool}. Noise results in variation in quality sampling which in turn leads to inconsistency between agents. As described in Section~\ref{sec:possibility}, applying the fusion function to inconsistent possibility distributions results in an increase in imprecision, i.e.\ to a fused possibility distribution with high possibility values for a number of different states. The resulting pignistic probability distribution (Definition~\ref{dfn:pignistic}) will then give higher probability to those states so that their quality values will be more likely to be re-sampled. In summary, we suggest that sampling from a noisy environment leads to more inconsistency between different agents' beliefs, and fusing those beliefs results in more imprecision which in turn results in more repeated sampling of a greater variety of states.

\subsection{Comparison with Probability}
In this subsection we directly compare the possibilistic model for best-of-$n$ with a probabilistic version. For the latter we adopt a discrete time model with the same structure as that of the possibilistic version but where agents' beliefs are represented by probability distributions which are then combined using the product fusion operator as introduced in Definition~\ref{dfn:probpool} below.\\
\begin{figure}[t]
    \centering
    \begin{subfigure}[b]{0.4\textwidth}
        \includegraphics[width=\textwidth]{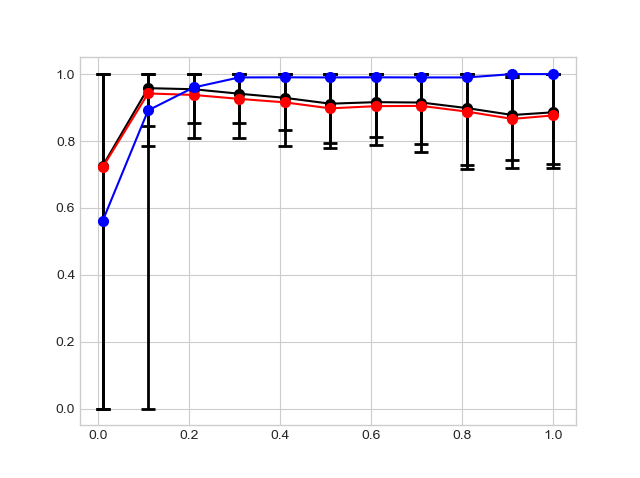}
        \put(-120,0){\footnotesize Evidence rate $\rho$}
        \caption{Comparison of possibility and probability for \\$\rho \in [0.01,1]$.}
        \label{fig:probposs}
    \end{subfigure}
    \begin{subfigure}[b]{0.4\textwidth}
        \includegraphics[width=\textwidth]{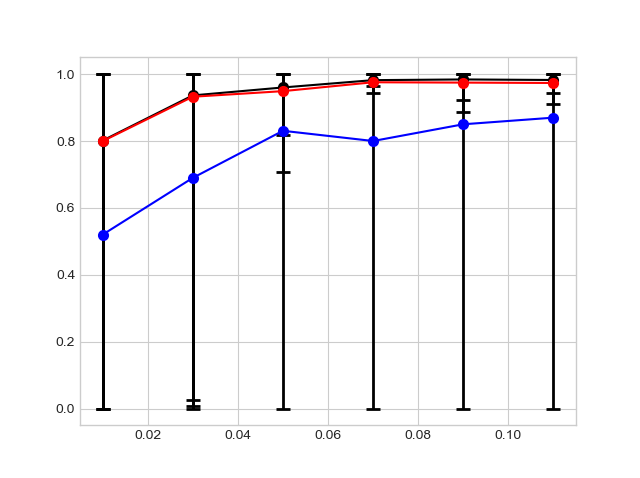}
        \put(-120,0){\footnotesize Evidence rate $\rho$}
        \caption{Comparison of possibility and probability for \\$\rho \in [0.01,0.11]$.}
        \label{fig:probpossshort}
    \end{subfigure}
    \caption{The average values of $\pi(s_n)$ (black line), $N(s_n)$ (red line) and probability $p(s_n)$ (blue line) for varying evidence rates $\rho$. These are after $1500$ iterations with $n=5$, $\theta=20$ and $\sigma=0.3$.}
    \label{fig:comparison}
\end{figure}
\begin{dfn}{Probability Fusion Function}\\
\label{dfn:probpool}
\textit{Let ${\mathbb P}$ denote the set of all probability distributions on ${\mathbb S}$. Then a probability fusion function is a function $c:{\mathbb P}^2 \rightarrow {\mathbb P}$. In particular, we will focus on the product fusion function given by: for $s_i \in {\mathbb S}$,
\begin{gather*}
c(p_1,p_2)(s_i)=\frac{p_1(s_i)p_2(s_i)}{\sum_{j=1}^n p_1(s_j)p_2(s_j)}.\\
\end{gather*}}
\end{dfn}
This is an established probability fusion operator~\cite{bordley} which as been studied for agent-based models combining fusion and evidential updating in~\cite{lee}, and in this context has been shown to have strong convergence properties. Figure~\ref{fig:comparison} compares the probabilistic and possibilistic models for $\sigma=0.3$ and for a range of evidence rates. For the probabilistic model agents are initialised with the uniform probability distribution, to represent ignorance, and for the possibilistic model we take Frank's parameter to be $\theta=20$ as above. For the probabilistic approach evidence from sampling the quality of state $s_i$ is represented by the following probability distribution:
\begin{align*}
p_E(s_j)=\begin{cases} \frac{1-q_i-\epsilon}{n} &: j \neq i,\\ \frac{(n-1)(q_i +\epsilon)+1}{n} &: j = i, \end{cases}
\end{align*} 
where $p_E$ corresponds to the linear combination of the probability distribution with $p(s_i)=1$ and the uniform distribution, with weights $q_i+\epsilon$ and $1-(q_i +\epsilon)$, respectively. On receiving this evidence an agent with current probability distribution $p$ updates to the distribution $c(p,p_E)$ where $c$ is the product fusion function given in Definition~\ref{dfn:probpool}.

\begin{figure}[t]
\centering
\includegraphics[width=0.4\textwidth]{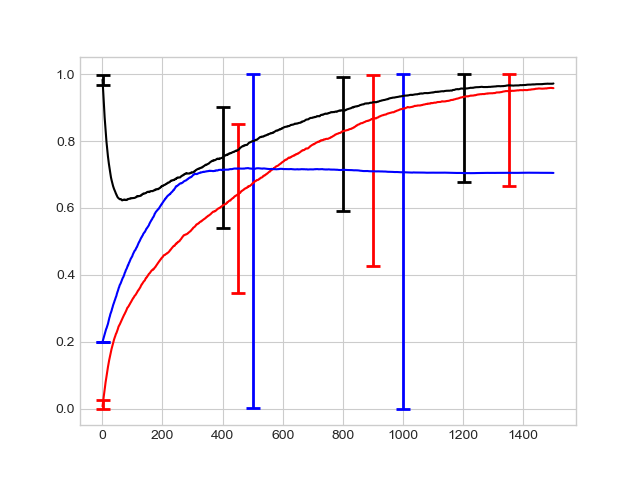}
\put(-110,0){\footnotesize Time $t$}
\caption{$\pi(s_n)$ (black line), $N(s_n)$ (red line) and $p(s_n)$ (blue line) plotted against time for $n=5$, $\theta=20$, $\rho=0.05$ and $\sigma=0.3$.}
\label{fig:probposstrajectory}
\end{figure}

\begin{figure}[b]
\centering
\includegraphics[width=0.4\textwidth]{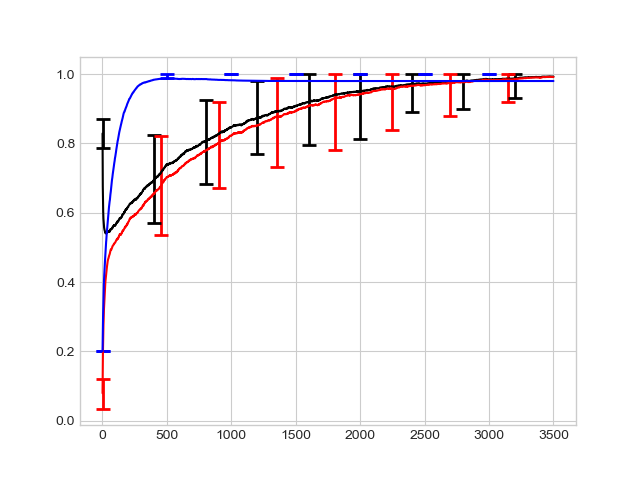}
\put(-110,0){\footnotesize Time $t$}
\caption{$\pi(s_n)$ (black line), $N(s_n)$ (red line) and $p(s_n)$ (blue line) plotted against time over $3500$ time steps, for $n=5$, $\theta=20$, $\rho=0.5$ and $\sigma=0.3$.}
\label{fig:probposstrajextra}
\end{figure}

Figure~\ref{fig:probposstrajectory} shows the average values of $\pi(s_n)$, $N(s_n)$ and also the average probability $p(s_n)$ obtained from the above probabilistic approach, plotted against time for $1500$ time steps when $\theta=20$, $\rho=0.05$ and $\sigma=0.3$. Clearly then, in such a scenario the possibilistic approach outperforms the probabilistic method in solving the best-of-$n$ problem under noise. However, the broader picture is somewhat more complex. For example, Figure~\ref{fig:comparison} shows the average values of $\pi(s_n)$, $N(s_n)$ and $p(s_n)$ at time step $t=1500$ for varying evidence rates. In particular, Figure~\ref{fig:probposs} shows these results for evidence rates between $0.01$ and $1$. These suggest that the possibilistic approach is more effective for lower evidence rates below $\rho=0.2$ (see Figure~\ref{fig:probpossshort}). However, the probabilistic approach performs better for higher evidence rates. While it is not entirely clear why this is the case we suggest that it might be related to the varying speeds of convergence of the two different approaches. For instance, Figure~\ref{fig:probposstrajextra} shows average possibility, necessity and probability values plotted against time for the higher evidence rate of $\rho=0.5$, but in this case for up to $3500$ time steps. Here we see that under possibilistic fusion and updating the population eventually converges to the same level of accuracy as obtained using the probabilistic approach but the former takes around $6$ times as long to converge as the latter. Given this, a possible explanation the difference in performance of the two approaches as summarised by Figure~\ref{fig:comparison} is that for very low evidence rates the probabilistic fusion operators converges before sufficient evidence has been received to ensure that the highest quality state is correctly identified.

\begin{figure}[t]
    \centering
    \begin{subfigure}{0.4\textwidth}
        \includegraphics[width=\textwidth]{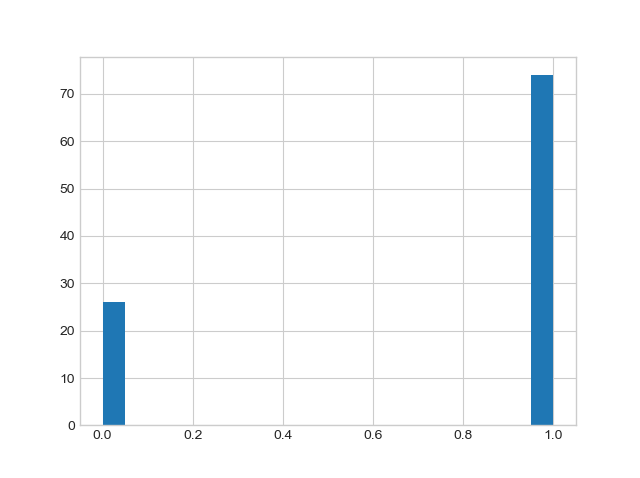}
         \put(-120,0){\footnotesize Average $p(s_n)$}
        \caption{Distribution of the average population value of $p(s_n)$ across $100$ runs.}
        \label{fig:histprob}
    \end{subfigure}
    \begin{subfigure}{0.4\textwidth}
        \includegraphics[width=\textwidth]{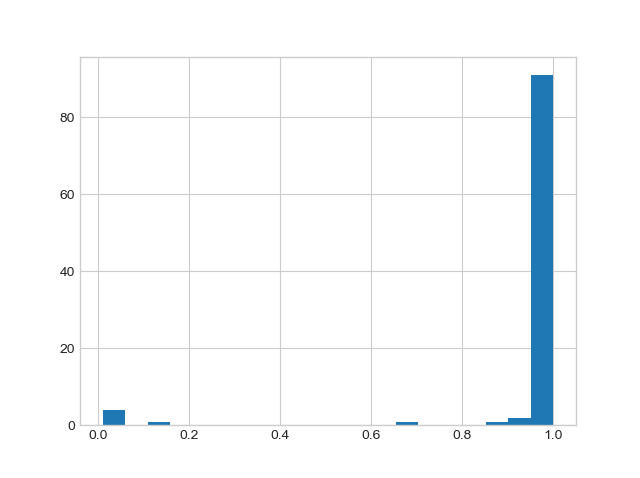}
        \put(-120,0){\footnotesize Average $\pi(s_n)$}
         \caption{Distribution of the average population value of $\pi(s_n)$ across $100$ runs.}
        \label{fig:histposs}
    \end{subfigure}
    \caption{Histograms of simulation outcomes at $t=1500$ across $100$ runs for both the probabilistic and possibilistic approaches when $\rho=0.05$ and $\sigma=0.3$.}
     \label{fig:histogram}
\end{figure}

This is to some extent confirmed by Figure~\ref{fig:histogram} which is a different presentation of the results shown in Figure~\ref{fig:probposstrajectory}. More specifically, Figures~\ref{fig:histprob} and \ref{fig:histposs} are respectively histograms of the average values of $p(s_n)$ and $\pi(s_n)$ across the $100$ independent runs of the simulation when $\sigma=0.3$ and $\rho=0.05$. For the probabilistic approach each run converges to consensus where all agents have probability values for $s_n$ either close to $1$ or close to $0$. Furthermore, there is significant proportion of runs in which the population reached consensus at $p(s_n)$ close to $0$. Note that the only stable fixed points of the probability fusion function given in Definition~\ref{dfn:probpool} are when both distributions give probability $1$ to the same state. Hence, there is an inherent tendency for a population of agents employing this fusion function to converge to one of these distribution. Evidence will then tend to skew convergence towards the correct distribution i.e.\ the distribution for which $p(s_n)=1$. However, in the case where there is very limited evidence and fast convergence then this skew may only be partial as we see in Figure~\ref{fig:histprob}. On the other hand, slower convergence when applying the possibility fusion function allows time, at least in the case of low evidence rates, for the distribution of outcomes across the different runs to be much more strongly skewed toward $\pi(s_n)=1$ as can be seen in Figure~\ref{fig:histposs}. However, there is then the disadvantage for the possibilistic approach in that, as the evidence rate increases, the convergence time also increases as more (partially conflicting) evidence becomes available at each time step, so that $1500$ time steps is no longer sufficient for the population to reach consensus on what is the best state as shown in Figure~\ref{fig:probposs}. Although given more time the population can converge on the correct answer as shown in Figure~\ref{fig:probposstrajextra}.

\section{Conclusions}
\label{sec:conclusions}
We have proposed an approach to distributed learning based on possibility theory, and shown that it can be effectively applied to the best-of-$n$ problem in a system where agents learn both directly from their environment and by fusing their beliefs with those of others. In particular, we have introduced a discrete time agent-based model and carried out a number of simulation experiments to study its meta-level properties and to gain insight into the performance of distributed possibilistic learning under varying evidence rates and levels of noise. 

Our results suggest that in a distributed learning context, possibility theory can provide a useful framework for representing uncertainty. This representation allows for the imprecision of an agent's belief to increase as a result of them encountering other beliefs which are partially inconsistent with their own. This in turn results in robust performance in the presence of noise, especially for low evidence rates for which the possibilistic approach outperforms a similar probabilistic model.

\section*{Acknowledgements}
This work was funded and delivered in partnership between the Thales Group, University of Bristol and with the support of the UK Engineering and Physical Sciences Research Council, ref.\ EP/R004757/1 entitled “Thales-Bristol Partnership in Hybrid Autonomous System Engineering (T-B PHASE).”


\end{document}